\documentclass[aps,prb,preprint,groupedaddress,showpacs]{revtex4-1}
\usepackage{graphicx}
\begin{document}

\title{Doping effect on the anomalous behavior of the Hall effect in electron-doped superconductor  Nd$_{2-x}$Ce$_x$CuO$_{4+\delta}$}

\author{T.B.Charikova$^{\rm 1}$, N.G.Shelushinina$^{\rm 1}$, G.I.Harus$^{\rm 1}$, D.S.Petukhov$^{\rm 1}$, A.V.Korolev$^{\rm 1}$, V.N.Neverov$^{\rm 1}$,
A.A.Ivanov$^{\rm 2}$}

\thanks{}
\affiliation{$^{\rm 1}$Institute of Metal Physics RAS, Ekaterinburg, Russia,
$^{\rm 2}$Moscow Engineering Physics Institute, Moscow, Russia}

\date{\today}

\begin{abstract}
Transport properties of Nd$_{2-x}$Ce$_x$CuO$_{4+\delta}$ single crystal films ($B \parallel {c}$,  $J \parallel {ab}$) are investigated in magnetic fields $B$ up to 9T at $T$=(0.4-4.2)K. An analysis of normal state (at $B>B_{c2}$) Hall coefficient $R_H$$^n$ dependence on Ce doping takes us to a conclusion about the existence both of electron-like and hole-like contributions to transport in nominally electron-doped system. In accordance with $R_H$$^n$(x) analysis an anomalous sign reversal of Hall effect in mixed state at $B<B_{c2}$ may be ascribed to a flux-flow regime for two types of carriers with opposite charges.
\end{abstract}


\maketitle

\section{Introduction}

An observation that Hall effect in the mixed state can have a sign opposite to that in the normal state has been reported for some conventional superconductor since 1960th. The discovery of a Hall effect sign change in the most of high-T$_c$ cuprate superconductors \cite{hagen,smith} stimulated a new interest to this anomalous behavior.

The high-T$_c$ superconductors are strongly type II and thus, in accordance with the theory of Abrikosov, in the mixed state their physical properties are determined by the presence of quantized flux lines or vortices. When an external current $\bf{J}$ is applied to the vortex system, the flux lines start to move under the action of the Lorentz force $\bf{F}$ = $\bf{J}$ $\times$ $\bf{B}$ with a velocity $\bf{v_L}$. As a consequence an electric field  $\bf{E}$ = -$\bf{v_L}$ $\times$ $\bf{B}$ is induced which has resistivity component $E_x$ and Hall component $E_y$, the longitudinal resistivity, $\rho_{xx}$ = $E_x$/$J$ and Hall resistivity $\rho_{xy}$ = $E_y$/$J$, thus appear.

Classical models which considered the hydrodynamical motion of vortices in response to a transport current in homogeneous structures (flux-flow models) were created at 1960th \cite{bardeen,nozieres}. In particular, Bardeen and Stephen (BS)\cite{bardeen} have  found that in the mixed state a part of the current flows through the vortex core that results in both dissipation and Hall effect. The core is treated as a normal conductor and thus in this model the Hall effect should have the same sign as in the normal state.

Various theoretical models have been proposed to explain the sign change of the Hall resistivity in the mixed state (for an extensive review see \cite{brandt}) but the origin of this phenomenon remains controversial.

If the proper pinning forces are included in the equation of vortex motion \cite{wang,jia} it leads to a back-flow current and to a possibility of Hall effect sign reversal. However, pinning forces are highly sample dependent and thus difficult to analyze.

Among the other models for the explanation of the Hall effect sign change in the mixed state a two-band model was created by Hirsh and Marsiglio (HM) \cite{hirsch} with electrons and holes supposed to have rather different superconducting gaps. Such a model will suggest a simple and natural explanation of the phenomenon but only if there are grounds to consider that two types of carriers, namely, electrons and holes, really coexist in the systems studied.

There are much recent activities in investigation of possible electrons and holes coexistence in the normal state of cuprate superconductors \cite{luo}. In particular, in electron-doped superconducting cuprates, angle resolved photoemission spectroscopy (ARPES) \cite{armitage1}$^-$\cite{ikeda2} and transport studies \cite{luo,fournier}$^-$\cite{charikova} have shown that both electrons and holes play a role in the normal state properties. Two kinds of carriers in electron-doped cuprates with different lanthanide cations seem to arise from the electronic structure near the Fermi surface (FS) of the CuO$_2$ planes. 

ARPES studies in NdCeCuO \cite {armitage2,matsui,matsui2} have revealed a small electron-like FS pocket in the underdoped region, and a simultaneous presence of both electron- and hole-like pockets near optimal doping. The conclusions of the ARPES measurements and first-principle calculations of the electronic structure on the electron-doped high-T$_c$ superconductors Ln$_{1.85}$Ce$_{0.15}$CuO$_4$ (Ln = Nd, Sm and Eu), performed by Ikeda et al. \cite{ikeda1,ikeda2} are in accordance with the results of papers  \cite {armitage2,matsui,matsui2}.

A spin density wave (SDW) model\cite{lin} was proposed which gives qualitative explanation to  ARPES observations. In this model, SDW ordering would induce FS reconstruction that results in an evolution from an electron pocket to the coexistence of electron-like and hole-like pockets and then into a single hole-like FS with increasing of doping. At present the development of this theoretical model is continued \cite{liu,kuchinskii}.

A model of FS reconstruction, induced by SDW, was equally well used for interpretation both of high-field Hall effect and magnetoresistance in electron-doped  Pr$_{2-x}$Ce$_x$CuO$_{4}$ films \cite{li} and of doping dependence of Shubnikov-de-Haas (SdH) oscillations in Nd$_{2-x}$Ce$_x$CuO$_{4}$ single crystals \cite{helm,kartsovnik}. A two-carrier model with coexisting of electrons and holes turns out to be suitable also for describing of temperature and doping dependencies of normal state Hall effect and thermopower in Nd$_{2-x}$Ce$_x$CuO$_{4}$ \cite{luo,fournier}, Pr$_{2-x}$Ce$_x$CuO$_{4}$ \cite{dagan} and  La$_{1.85}$Ce$_{0.15}$CuO$_4$  thin films \cite{jin,zhao}. 

Thus, it seems to us that it is  just the time to make a new attempt of describing the behavior of the mixed state Hall effect in electron-doped superconductors on the ground of a two-carrier model. In this paper we study magnetic field dependencies of longitudinal and Hall resistivities of electron-doped superconductor Nd$_{2-x}$Ce$_x$CuO$_{4+\delta}$ with various Ce content both in the normal and the mixed states.

\section{Samples and equipment}

The series of Nd$_{2-x}$Ce$_x$CuO$_{4+\delta}$ epitaxial films ($x$ =  0.14, 0.15, 0.17, 0.18) with standard (001) orientation were synthesized by pulsed laser deposition \cite{charikova}. Then the films were subjected to heat treatment (annealing) under various conditions to obtain samples with various oxygen content.  As a result, three types of samples with $x$=0.15 were obtained: as-grown samples, optimally reduced samples (optimally annealed in a vacuum at T = 780$^0$\,C for t = 60 min; p = 10$^{-2}$\,mmHg) and non optimally reduced samples (annealed in a vacuum T = 780$^0$\,C for t = 40 min; p = 10$^{-2}$\,mmHg), five types of samples with $x$=0.14: optimally annealed in a vacuum at T = 780$^0$\,C for t = 25 min; p = 10$^{-2}$\,mmHg and four non optimally reduced samples (annealed in a vacuum T = 780$^0$\,C for t = 5, 20, 30, 64 min; p = 10$^{-2}$\,mmHg), ten types of samples with $x$ = 0.17: optimally annealed in a vacuum at T = 780$^0$\,C for t = 50 min; p = 10$^{-2}$\,mmHg) and non optimally reduced samples (annealed in a vacuum T = 780$^0$\,C for different times; p = 10$^{-2}$\,mmHg) and five types of samples with $x$=0.18: optimally annealed in a vacuum at T = 600$^0$\,C for t = 35 min; p = 10$^{-5}$\,mmHg and four non optimally reduced samples (annealed in a vacuum T = 600$^0$\,C for t = 10, 15, 25,60 min; p = 10$^{-5}$\,mmHg) The film thickness was 1200 - 3800 \AA. 

A part of experimental results on the galvanomagnetic properties of these films are presented earlier \cite{charikova2}. Now we report on the investigation results of the magnetic field dependent Hall coefficient as a function of doped electron concentration in optimally reduced films. Hall effect and magnetoresistance were measured by a standard dc technique using a 12T Oxford Instruments superconducting magnet in the temperature range $T =$ (0.4 $\div$ 4.2) K.

\section{Experimental results and discussion}

The in-plane longitudinal $\rho_{xx}$ and Hall $\rho_{xy}$ components of the resistivity are measured as the functions of  perpendicular to the ab-plane magnetic field $B$ up to 9T in single crystal films of electron-doped superconductor Nd$_{2-x}$Ce$_x$CuO$_{4+\delta}$ at $T$=(0.4 - 4.2)K. Fig.1 shows the field dependences of the resistivities $\rho_{xx}$ and $\rho_{xy}$ for optimally reduced Nd$_{2-x}$Ce$_x$CuO$_{4+\delta}$  films with $x$ = 0.14 (a) and 0.15 (b) and Fig.2 shows $\rho_{xx}(B)$ and $\rho_{xy}(B)$ dependences for optimally reduced Nd$_{2-x}$Ce$_x$CuO$_{4+\delta}$  films with  $x$ = 0.17 (c) and 0.18 (d) for $T$ = 4.2 K. Here $B_p$ is a vortex-depinning field and $B_{c2}$ is the upper critical field. The region  $B_p < B< B_{c2}$  corresponds to a mixed (vortex) state where finite resistivity is a consequence of vortex moving under the action of the Lorentz force.

\begin{figure}
\includegraphics{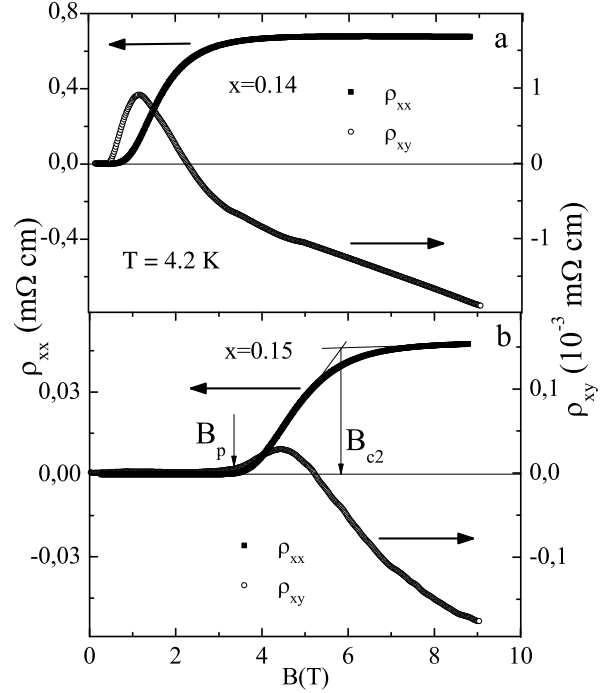}
\caption{\label{fig:wide}In-plane resistivity $\rho_{xx}$ and Hall resistivity $\rho_{xy}$ versus   magnetic field $B$ plots for optimally annealed single crystal films of  Nd$_{2-x}$Ce$_x$CuO$_{4+\delta}$ with $x$ = 0.14 (a) and 0.15 (b). $B_p$ is a vortex-depinning field and $B_{c2}$ is the upper critical field.}
\end{figure}

The evolution of the Hall coefficient value in the normal state above the upper critical field $B_{c2}$ is traced with a variation of Ce doping. It is found that low temperature normal state Hall coefficient $R_H^n$ is negative for underdoped ($x$=0.14) and optimally doped ($x$=0.15) films, positive for highly overdoped ($x$=0.18) films and has $R_H^n \cong$ 0 for slightly overdoped ($x$=0.17) films in accordance with previous results for normal state $R_H$  of Nd$_{2-x}$Ce$_x$CuO$_{4+\delta}$ at $T > T_c$  \cite{uchida}$^-$\cite{xu} (Fig. 3).

In other electron-doped superconductors the situation is the same: in Pr$_{2-x}$Ce$_x$CuO$_{4}$  $R_H$$^n <$ 0 for $x$ = (0.11 - 0.15) and $R_H$$^n >$ 0 for $x$ = (0.16 - 0.19) at 0.35 K and $B$ = 14 T \cite{armitage2}$^,$\cite{dagan}; in La$_{2-x}$Ce$_x$CuO$_4$  the sign of the Hall resistivity changes from negative for $x$ = (0.06-0.105) to positive for $x$ = (0.12-0.17) at 2 K  and $B$ = 10 T \cite{jin,zhao}.

\begin{figure}
\includegraphics{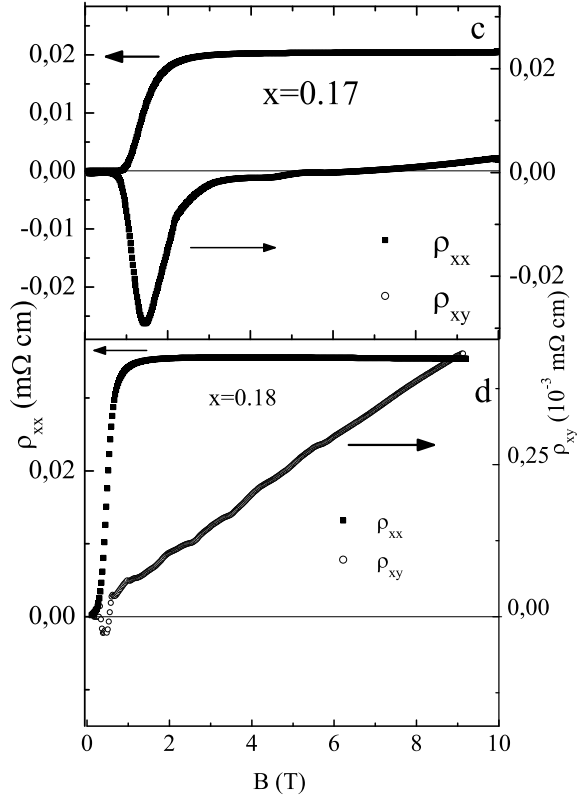}
\caption{\label{fig:wide}In-plane resistivity $\rho_{xx}$ and Hall resistivity $\rho_{xy}$ versus   magnetic field $B$ plots for optimally annealed single crystal films of  Nd$_{2-x}$Ce$_x$CuO$_{4+\delta}$ with $x$ = 0.17 (c) and 0.18 (d).}
\end{figure}

\begin{figure}
\includegraphics{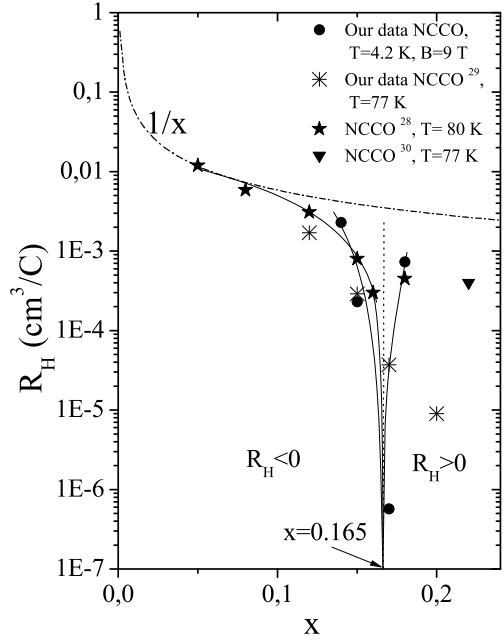}
\caption{\label{fig:wide}Hall coefficient $R_H$ in the normal state of Nd$_{2-x}$Ce$_x$CuO$_{4+\delta}$ (NCCO) systems as a function of Ce content. The  dash-doted theoretical curve corresponds to the one-carrier model. The solid circles are our experimental data, solid lines are guides for eyes.}
\end{figure}

Such a behavior of  $R_H^n$(x) may be naturally explained by a coexistence of electrons (of concentration $n$) with holes (of concentration $p$) even in nominally electron-doped cuprate system. Proceeding from a relation $n \sim x$ for ideal electron doping Luo \cite{luo} carried out a division of electron and hole contributions and came to a conclusion about rather rapid increase of $p$($x$). The two types of carriers in cuprates may originate from electron-like and hole-like parts of the Fermi surface in agreement with the ARPES results \cite{armitage2}.

As in the most of cuprates we have found an anomalous change  of the Hall effect sign in the mixed state $R_H$$^f$ just below $B_{c2}$: from negative to positive for $x$=0.14; 0.15 and from positive to negative for $x$=0.18 (see Figs.1,2). This Hall anomaly has been previously interpreted using a lot of different assumptions, most of which were closely connected with the peculiarities of magnetic flux dynamics in the mixed state of superconductors \cite{brandt}$^-$\cite{jia,feigelman,khomskii}.

We have adopted a semi phenomenological description of a mixed state Hall effect by flux-flow model of Bardeen and Stephen \cite{bardeen} modified by coexistence of electrons and holes. In analogy with HM \cite{hirsch} we have proposed that two types of carriers have rather different superconducting gaps and thus different upper critical fields $B_{c2}$$^{(i)}$, $i$ = 1, 2 for electrons and holes, respectively.
Really, taking into account of an antiferromagnetic spin fluctuations leads naturally to a two-band/two-gap model in superconducting state of the electron-doped cuprate \cite{liu}.

Magnetic field dependencies of resistivity $\rho_{xx}(B)$ and Hall coefficient $R_H(B)$, usually observed in semiconductors with two (or several) types of carriers \cite{blatt}, are associated both with strong-field condition $\omega_c \tau \gg $ 1 ($\omega_c$ being a cyclotron frequency and $\tau$ being an momentum relaxation time) at least for one of the types and with widely different carrier mobilities.

In contrast, in our situation magnetic field dependencies of $\rho_{xx}$ and $R_H$, even with $\omega_c \tau \ll $ 1 for each types of carriers, are caused by conditions of the flux-flow regime and the interplay between two types in a magnetic field arises due to widely different rates of their transition into the normal state (different $B_{c2}^{(i)}$).

In the framework of a conventional Drude model the low-field Hall coefficient for two types of carriers with partial conductivities $\sigma_i$ and Hall coefficients $R_H^{(i)} \equiv R_i$ ($i$  = 1, 2) is given by \cite{blatt}

\begin{equation}
R_H =\frac{\sigma_1^2 R_1 + \sigma_2^2 R_2}{\sigma^2},
\end{equation}

where the total conductivity $\sigma = \sigma_1 + \sigma_2$. For electrons ($i$ = 1) with concentration $n$ and mobility $\mu_n$ and holes ($i$ =2) with concentration $p$ and mobility $\mu_p$ in the normal state we have:
\begin{eqnarray}
\sigma_{01} = e n \mu_n;   \enspace \sigma_{02} = e p \mu_p
\nonumber\\
R_{01} = - \frac {1}{e n};   \enspace  R_{02} = \frac {1}{e p} 
\end{eqnarray}

In the HM paper\cite{hirsch} the equation for the longitudinal resistivity $\rho_{xx}$ in the mixed state was taken as in the BS model \cite{bardeen}, but in the equation for a Hall angle it was supposed  that the ratio of the normal part of electrons (holes) in the mixed state to the total number of electrons (holes) is proportional to $B/B_{c2}^{(i)}$. In contrast to that we use the standard BS dependencies on $B$  both of $\rho_{xx}$ and $\rho_{xy}$  for each of the types of carriers in the mixed state:

\begin{eqnarray}
\rho_{xx}^f = \rho_{xx}^0 \cdot \frac{B}{B_{c2}},
\nonumber\\
\nonumber\\
\rho_{xy}^f = \omega_c \tau \rho_{xx}^0 \frac {B}{B_{c2}} = \rho_{xy}^0 \cdot \frac{B}{B_{c2}}, 
\end{eqnarray}

where $\rho_{xx}^0 = 1/\sigma_0$ and $\rho_{xy}^0 = R_0 \cdot B$ are the normal state resistivities and $R_0$ is the normal state Hall coefficient.

As in the low-field limit $\sigma^f = 1/\rho_{xx}^f$ and $R^f = \rho_{xy}^f/B$, thus we propose:

\begin{equation}
\sigma_{i}^f = \sigma_{0i} \cdot \frac{B_i}{B}, \enspace  R_i^f = R_{0i} \cdot \frac{B}{B_i}, 
\end{equation}

where for simplicity $B_i \equiv B_{c2}^{(i)}$. 

Then we have three regions of magnetic field:

(a) $B < B_1$, where both of electrons and holes are in the mixed state;

(b) $B_1 < B < B_2$, where electrons are in the normal state but holes are in the mixed one;

(c) $B > B_2$, where both electrons and holes are in the normal state.

From Eqs. (1), (2) and (4) after some algebra we obtain simple formulas, suitable for desciption of experimental dependencies ($\rho_{xx} = 1/\sigma$):

\begin{eqnarray}
(a) \qquad  \frac{R}{\mid R_{0n}\mid} = \frac{\gamma b^2 \frac{B_2}{B_1} - 1}{(\gamma b \frac{B_2}{B_1} + 1)^2} \cdot \frac{B}{B_1}, 
\nonumber\\
\nonumber\\
\frac{\sigma}{\sigma_{0n}} = (1 + \gamma b \frac{B_2}{B_1}) \cdot \frac{B_1}{B};
\end{eqnarray}

\begin{eqnarray}
(b) \qquad  \frac{R}{\mid R_{0n}\mid} = \frac{\gamma b^2 \frac{B_2}{B} - 1}{(\gamma b \frac{B_2}{B} + 1)^2}, 
\nonumber\\
\nonumber\\
\frac{\sigma}{\sigma_{0n}} = 1 + \gamma b \frac{B_2}{B};
\end{eqnarray}

\begin{eqnarray}
(c) \qquad  \frac{R}{\mid R_{0n}\mid} = \frac{\gamma b^2 - 1}{((\gamma b + 1)^2}; 
\nonumber\\
\nonumber\\
\frac{\sigma}{\sigma_{0n}} = 1 + \gamma b;
\end{eqnarray}

where $\mid R_{0n}\mid \equiv 1/en$, $\sigma_{0n} = en \mu_n$, $\gamma = p/n$ and $b = \mu_p /\mu_n$.

Fig.4 shows a comparison of experimental dependencies of  $R_H$ with ones calculated by means of formulas (5)-(7) for $x$ = 0.14 (a) and $x$ = 0.15 (b) films. An effect of pinning is taken into account phenomenologically by a counting of field $B$ in formulas (5)-(7) from the depinning field $B_p$ and by a supposition that at $B < B_p$ $\rho_{xx}$ = 0 and $R_H$ = 0.

As it is followed from Eqs.(5)-(7) a field of anomalous peak maximum $B_{max} = B_1$ and experimental value of the upper critical field $B_{c2}^{exp} = B_2$. Thus, only carrier concentrations $n, p$ and mobilities $\mu_n, \mu_p$ come out as the fitting parameters. The values of parameters obtained by fitting procedure both for $\rho_{xx}(B)$ and for $R_H(B)$ are presented in Table 1.

\begin{figure}
\includegraphics{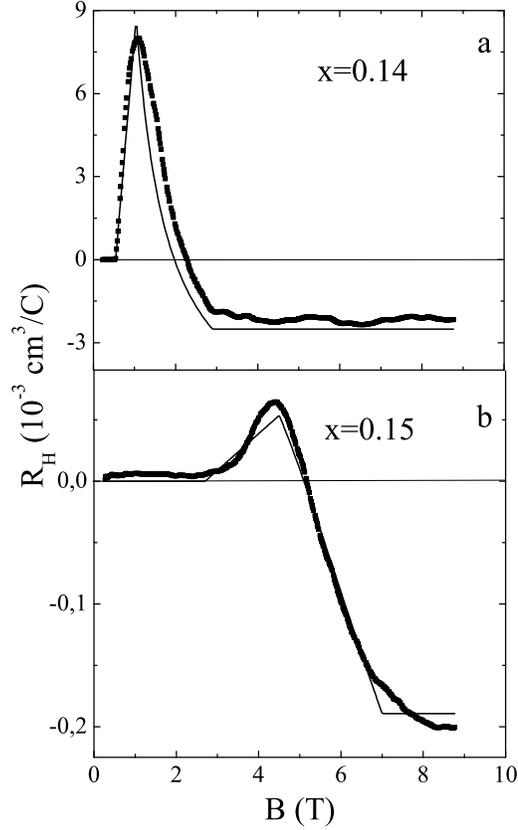}
\caption{\label{fig:wide}The experimental behavior of $R_H$  as a function of magnetic field (points) in electron-doped superconductor Nd$_{2-x}$Ce$_x$CuO$_{4+\delta}$ with $x$ = 0.14 and 0.15 and calculated $R_H$($B$) dependences (solid lines) for reasonable values of parameters (see Table 1).}
\end{figure}

We see that it turned out possible to describe qualitatively the behavior of $R_H(B)$  both in the normal and mixed states of Nd$_{2-x}$Ce$_x$CuO$_{4+\delta}$ films with $x$ = 0.14 and 0.15 for $b \approx 1$ and reasonable values of parameters $n$ and $p$: for a small magnetic fields $B_p < B < B_1$ (region (a)) the electrons will become ``normal'' more rapidly than the holes ($B_1 < B_2$) and thus the Hall coefficient have a good chance to be positive, switching to negative values at $B = B_{inv}$ in a region (b) and then tends to its normal (negative) value in a region (c). In particular, a field of $R_H$ sign inversion is given by $B_{inv} = \gamma b \cdot B_2$.

\begin{figure}
\includegraphics{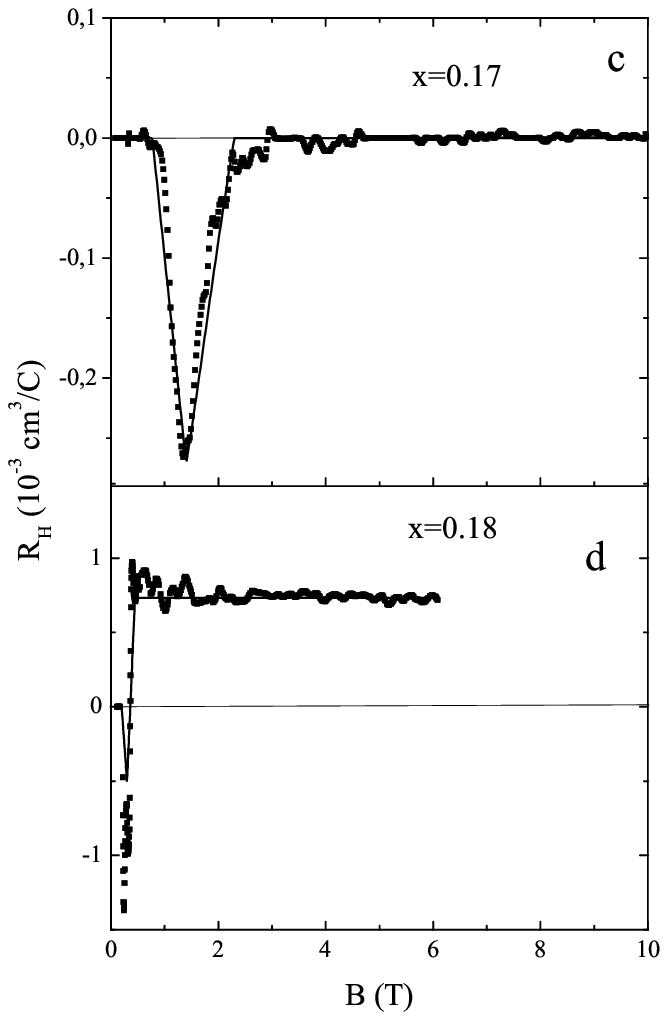}
\caption{\label{fig:wide}The experimental behavior of $R_H$  as a function of magnetic field in electron-doped superconductor Nd$_{2-x}$Ce$_x$CuO$_{4+\delta}$ with $x$ = 0.17 and 0.18 (points) and calculated $R_H$($B$) dependences (solid lines) for reasonable values of parameters (see Table 1).}
\end{figure}

\begin{table}[b]
\caption{\label{tab:table1}
The values of  parameters obtained from fitting of Eqs. (5)-(7) to the experimental data on $\rho_{xx}(B)$ and $R_H(B)$.}
\begin{ruledtabular}
\begin{tabular}{ccccccc}
$x$ &$n$ &$p$ & $\gamma$& $\mu_n$ & $\mu_p$ & $b$   \\
    & (cm$^{-3}$)& (cm$^{-3}$)&  & ($\frac{cm^2}{B c}$)& ($\frac{cm^2}{B c}$)& \\
\hline
0.14&  6.9$\cdot$10$^{20}$ & 1.04$\cdot$10$^{20}$ & 0.15 & 10 & 20& 2 \\
0.15&  4.3$\cdot$10$^{21}$  & 5.4$\cdot$10$^{21}$  & 1.3 & 18& 12& 0.6 \\
0.17&  1.7$\cdot$10$^{21}$  & 6.7$\cdot$10$^{21}$  & 4.0 & 60& 30& 0.5 \\
0.18&  1.6$\cdot$10$^{19}$  & 1.2$\cdot$10$^{21}$  & 79.5 & 100 & 90& 0.9 \\
\end{tabular}
\end{ruledtabular}
\end{table}

For overdoped systems where Hall coefficient is positive in the normal state and negative in the mixed state (see Fig.2) in formulas, analogous to (5)-(7), for reasonable description of experimental data it should be taken $B_2 < B_1$ and thus $B_{max} = B_2$, $B_{c2}^{exp} = B_1$ and $B_{inv} = (\gamma b)^{-1} \cdot B_1$.

In Table 1 the parameters $n, p$ and $\mu_n, \mu_p$ obtained from fitting of calculated dependencies of $\rho_{xx}(B)$ and $R_H(B)$ to the experimental ones for Nd$_{2-x}$Ce$_x$CuO$_{4+\delta}$  with $x$ = 0.17 and 0.18 (Fig.5) are also presented.

It is seen that we obtained $b \cong$ 1 for all the samples and $n >> p$ for $x$ = 0.14, $n \cong p$ for $x$ = 0.15, $p > n$ for $x$ = 0.17 and $p >> n$ for $x$ = 0.18. Such a trend is in qualitative accordance with estimates from $R_H^n(x)$ dependencies \cite{luo}, from ARPES data \cite{armitage2}, from frequencies of SdH oscillations \cite{helm,kartsovnik} for different Nd$_{2-x}$Ce$_x$CuO$_{4}$ samples and from high-field $\rho_{xy}(B)$ dependencies for Pr$_{2-x}$Ce$_x$CuO$_{4}$ systems\cite{li}.

Thus, an anomalous sign reversal of Hall effect in the mixed state of investigated systems may be described rather well by a flux-flow model for two types of carriers with opposite charges.

\section{Conclusions}

We have analysed the Ce doping dependence of the normal state Hall coefficient in optimally reduced Nd$_{2-x}$Ce$_x$CuO$_{4+\delta}$ single crystal films and, on the grounds of this analysis, have recruited a two-carrier model for describing of the mixed state Hall coefficient. Our scheme is based on a simple Drude model for normal state and on a semiphenomenological Bardeen-Stephen model for the mixed one modified by coexistence of electrons and holes. 

This description may be considered as an illustration of a possibility for the natural explanation of Hall effect sign reversal systematically observed in a mixed state of cuprate superconductors. Such an approach may occur actual in the light of much recent efforts on experimental (ARPES) and theoretical (SDW-model) investigations of the electronic structure near Fermi surface in the CuO$_2$ plane of high-T$_c$ superconducting systems.

\begin{acknowledgments}
This work was done within RAS Program (project N 09-P-2-1005 Ural Division RAS) with partial support of RFBR (grant N 09-02-96518).
\end{acknowledgments}

\end{document}